\documentclass[prd,aps,showpacs,preprintnumbers,amssymb]{revtex4}
\usepackage{color}
\usepackage{axodraw}
\usepackage{epsf}

\def\e3p{$\eta \rightarrow 3 \pi$}

\begin{document}

\title{%
\hfill{\normalsize\vbox{%

 }}\\
{Higgs boson otherwise}}

\author{Renata Jora $^{\it \bf b}$~\footnote[1]{Email:
 cjora@physics.syr.edu}}
\affiliation{$^ {\bf \it b}$ INFN Roma, Piazzale A Moro 2,
Roma, I-00185 Italy.}

\date{\today}

\begin{abstract}
 We propose an electroweak model based on the identification of the Higgs
 with the dilaton. We show that it is possible in this context to have a massless Higgs boson at tree and one loop levels without contradicting the main experimental and theoretical constraints.
\end{abstract}


\maketitle
\vspace{1cm}
\section{Introduction}
 The elusive Higgs boson and its mass are as yet to be found at
  high energy colliders like Tevatron and LHC.
  There are known theoretical constraints \cite{Kane1}, \cite{Kane2} coming from unitarity,
  triviality and vacuum stability which set bounds on it,
  but in general they are not as strong as
 the experimental ones.
 Thus LEP sets the lower limit at $114.4 $ GeV and the electroweak precision date place the Higgs boson at $76^{+111}_{-38}$.

 Throughout the years very important insights and theorems
 about the behavior of various types of quantum field theories have been obtained using the scale invariance.
  Associated with it is the dilaton with roots in our case in the technicolor theories.
   In this paper we propose an electroweak model where the role of the Higgs boson is played by the dilaton.
  While models of this type have been analyzed before \cite{Skiba1},\cite{Skiba2}
   here we introduce a specific representation of the Higgs inspired by QCD \cite{JS} along with a particular Higgs potential. Thus we propose a situation in which although the scale invariance is broken at tree level or at one loop level the Higgs boson is massless.

   Of course, even in these circumstances when the couplings of the Higgs are very different than those of the SM it remains for us to explain how it is possible that a completely massless Higgs boson has not yet been detected and how we can avoid various experimental and theoretical constraints. We postpone that for section VI where we also suggest possible collider signatures.

In section II we will briefly mention the one loop effective potential in
 the standard model.
 Since this is a well known subject the details can be found elsewhere \cite{Sher1},\cite{MV}.

 In section III we discuss the Higgs boson self energy. In section IV we give a brief account of the quadratic divergences. In section V we propose a Higgs potential which can lead to a massless Higgs boson and analyze the consequences of this choice.
Section VI and VII are dedicated to phenomenology and respectively, conclusions.

\section{The model}

 We consider an electroweak model invariant under the larger group $SU(2)\times SU(2)$ \cite{TA} and we gauge only one of them. Thus the Higgs fields will be given by a $2\times2$ matrix:
\begin{eqnarray}
\Phi=f\exp [I\frac{1}{f} \left[
\begin{array}{ccc}
\pi_0-I\sigma&\pi_1-I\pi_2\\
\pi_1+I\pi_2&-\pi_0-I\sigma
\end{array}
\right] ]
\label{def45637}
\end{eqnarray}

 Then we can break the scale invariance and one of the $SU(2)$ at the same scale f. This would give vev different than zero for only one field since the invariant potential would depend only on one of them as follows:

\begin{eqnarray}
Tr(\Phi\Phi^{\dagger})=Tr(f\exp [I \frac{1}{f}\left[
\begin{array}{ccc}
\pi_0-I\sigma&\pi_1-I\pi_2\\
\pi_1+I\pi_2&-\pi_0-I\sigma
\end{array}
\right] ]
f \exp [-I\frac{1}{f} \left[
\begin{array}{ccc}
\pi_0+I\sigma&\pi_1+I\pi_2\\
\pi_1-I\pi_2&-\pi_0
+I\sigma
\end{array}
\right]]
)
=f \exp{2\frac{\sigma}{f}}
\label{matrix11}
\end{eqnarray}

We will take the dilaton potential at tree level as:
\begin{equation}
V=a_1(\chi-f)^4
\label{newdilpot}
\end{equation}
where $\chi=fe^{\frac{\sigma}{f}}$.

This potential breaks explicitly the scale invariance and gives the vev for the dilaton field $\chi=f$. It is evident that the mass of the dilaton field calculated at the minimum of the potential is zero. Thus at tree level the dilaton is exactly massless, although the scale invariance is broken. If instead of one introduces matrix representation of $SU(2)$ then all it is to do is to rewrite Eq (\ref{newdilpot}) as:
\begin{equation}
V=a_1[Tr(\Phi^{\dagger}\Phi)^{\frac{1}{2}}-f]^4
\label{newdilpot1}
\end{equation}

First note that a square root in the potential does not affect at all the invariance and that from the set up of the model neither the renormalizability.

In order to agree with the established results the relevant coupling constants in the model are rescaled as follows:
\begin{eqnarray}
&&g\rightarrow \frac{v}{f}g
\nonumber\\
&&g'\rightarrow\frac{v}{f}g'
\nonumber\\
&&g_{y}\rightarrow \frac{v}{f}g_{Y}
\label{scaling}
\end{eqnarray}

\section {The effective Higgs potential}

The tree level Higgs potential in  the standard model has the form:
\begin{equation}
U(\phi)=\frac{1}{2}\mu^2\Phi^{\dagger}\Phi+\frac{1}{4}\lambda
(\Phi^{\dagger}\Phi)^2 \label{thepotential}
\end{equation}
We consider here the most general form of the potential not necessarily renormalizable.
Since the scalar part of the potential has been calculated in detail
in \cite{CW}, \cite{W} and more recently in \cite{Sher1} we will
give here only the main results. We are working in the Landau gauge
where there are no coupling between the Goldstone bosons and the
physical Higgs.

Mainly the one loop contributions coming from the scalars is obtained from the diagrams in Fig. 1.
\vspace{0.5cm}

\begin{center}
\SetScale{2.5}
\begin{picture}(250,100)(0,0)
\GCirc(20,20){5}{1}
\Line(20,15)(20,12)
\Line(20,15)(21.5,12.4)
\Line(20,15)(18.5,12.4)
\Text(87.5,50)[]{$+$}
\GCirc(50,20){5}{1}
\Line(50,15)(50,12)
\Line(50,15)(51.5,12.4)
\Line(50,15)(48.5,12.4)
\Line(50,25)(50,28)
\Line(50,25)(51.5,27.4)
\Line(50,25)(48.5,27.4)
\Text(162.5,50)[]{$+$}
\GCirc(80,20){5}{1}
\Line(80,15)(80,12)
\Line(80,15)(81.5,12.4)
\Line(80,15)(78.5,12.4)
\Line(84.33,22.5)(87.33,22.5)
\Line(84.33,22.5)(86.92,24)
\Line(84.33,22.5)(85.83,25.09)
\Line(75.67,22.5)(72.67,22.5)
\Line(75.67,22.5)(73.07,24)
\Line(75.67,22.5)(74.17,25.09)
\Text(236.5,50)[l]{$+...$}
\end{picture}
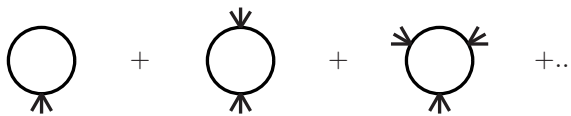
\\{\sl Fig. 1. Diagrams relevant for the scalar interaction.}
\end{center}

The one loop contribution coming from the usual Higgs is:
\begin{equation}
V_1(\phi)=\frac{1}{2}\int \frac {d^4
k}{(2\pi)^4}\ln(1+\frac{\frac{\partial^2 U}{\partial h^2}
(\phi)}{k^2}) \label{onelooppotential1}
\end{equation}

The one loop contribution coming form the three Goldstone bosons has
the expression:
\begin{equation}
V_2(\phi)=\frac{1}{2}\int \frac {d^4
k}{(2\pi)^4}\ln(1+\frac{\frac{\partial^2 U}{\partial G^2}
(\phi)}{k^2}) \label{onelooppotential2}
\end{equation}

In general the one loop correction to the potential can be read off from the tadpole diagrams given in Fig.2.

\vspace{0.5cm}
\begin{center}
\begin{picture}(240,30)(0,0)
\DashCArc(20,20)(10,0,360){1}
\GCirc(90,20){10}{1}
\PhotonArc(160,20)(10,0,360){1}{8}
\GlueArc(230,20)(10,0,360){1}{8}
\DashLine(20,10)(20,0){2}
\DashLine(90,10)(90,0){2}
\DashLine(160,10)(160,0){2}
\DashLine(230,10)(230,0){2}
\end{picture}
\vspace{0.5cm}
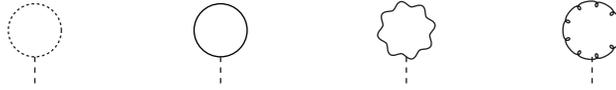
\\{\sl Fig. 2. Tadpole diagrams in order for the scalar, spinor field, gauge field and ghosts.}
\end{center}

 Introducing a cut-off $\Lambda$ the expressions in Eq(\ref{onelooppotential1})-(\ref{onelooppotential2}) reduce to:

 \begin{equation}
V_1(\phi)=\frac{1}{16\pi^2}[
\frac{1}{4}U_h''^2(\ln\frac{U_h''}{\Lambda^2}-\frac{1}{2})
+\frac{1}{2}\Lambda^2U_h'']
\label{result1}
\end{equation}

 \begin{equation}
V_2(\phi)=3\frac{1}{16\pi^2}[
\frac{1}{4}U_G''^2(\ln\frac{U_G''}{\Lambda^2}-\frac{1}{2})
+\frac{1}{2}\Lambda^2U_G'']
\label{result1}
\end{equation}

Here we denote $U_h''=\frac{\partial^2 U}{\partial h^2}$ and
$U_G''=\frac{\partial^2 U}{\partial G^2}$.

 The contribution from the gauge bosons can be calculated in a similar way:

\begin{eqnarray}
V_3(\phi)&=&\frac{3}{64\pi^2}[\frac{1}{16}g^4\phi^4(\ln\frac
{g^2\phi^2}{4\Lambda^2}-\frac{1}{2}) +\frac{1}{2}\Lambda^2 g^2\phi^2
\nonumber\\
V_4(\phi)&=&\frac{3}{64\pi^2}[\frac{1}{16}(g^2+g'^2)^2\phi^4(\ln\frac
{(g^2+g'^2)\phi^2}{4\Lambda^2}-\frac{1}{2}) \nonumber\\ &+&\frac{1}{2}\Lambda^2
(g^2+g'^2)\phi^2
]
\label{result2}
\end{eqnarray}

 Among the fermions we consider only the contribution of the
 top quark since this is the most significant:

\begin{equation}
 V_5(\phi)
 =-\frac{12}{64\pi^2}[\frac{1}{4}g_Y^4\phi^4(\ln\frac
{g_Y^2\phi^2}{2\Lambda^2}-\frac{1}{2}) +\Lambda^2 g_Y^2\phi^2
]
\label{results3}
\end{equation}

In general the one loop potential is the sum of the above terms:

\begin{equation}
V(\phi)=U(\phi)+V_1+V_2+V_3+V_4+V_5
\label{fullpotential}
\end{equation}

In our case of interest however the term corresponding to the Goldstone bosons does not appear.

\section{The scalar self energy}

We work in the Landau gauge where the Goldstone bosons are decoupled from the usual Higgs.
The scalar self energy is determined from the diagrams in Fig. 3.
\vspace{0.5cm}
\begin{center}
\begin{picture}(250,150)(0,0)
\DashCArc(20,20)(10,0,360){1}
\GCirc(90,20){10}{1}
\PhotonArc(160,20)(10,0,360){1}{8}
\GlueArc(230,20)(10,0,360){1}{8}
\DashLine(20,10)(20,0){2}
\DashLine(5,0)(35,0){2}
\DashLine(90,10)(90,0){2}
\DashLine(75,0)(105,0){2}
\DashLine(160,10)(160,0){2}
\DashLine(145,0)(175,0){2}
\DashLine(230,10)(230,0){2}
\DashLine(215,0)(245,0){2}
\DashLine(5,60)(35,60){2}
\DashLine(75,60)(105,60){2}
\DashLine(145,70)(150,70){2}
\DashLine(170,70)(175,70){2}
\DashLine(215,70)(220,70){2}
\DashLine(240,70)(245,70){2}
\DashCArc(20,70)(10,0,360){1}
\PhotonArc(90,70)(10,0,360){1}{8}
\GCirc(160,70){10}{1}
\GlueArc(230,70)(10,0,360){1}{8}
\DashLine(15,110)(22.5,110){2}
\DashLine(42.5,110)(50,110){2}
\DashLine(102.5,110)(110,110){2}
\DashLine(130,110)(137.5,110){2}
\DashLine(190,110)(197.5,110){2}
\DashLine(217.5,110)(225,110){2}
\PhotonArc(32.5,110)(10,0,360){1}{8}
\DashCArc(120,110)(10,0,360){1}
\DashCArc(207,110)(10,0,180){1}
\PhotonArc(207,110)(10,180,360){1}{4}
\end{picture}
\vspace{0.5cm}
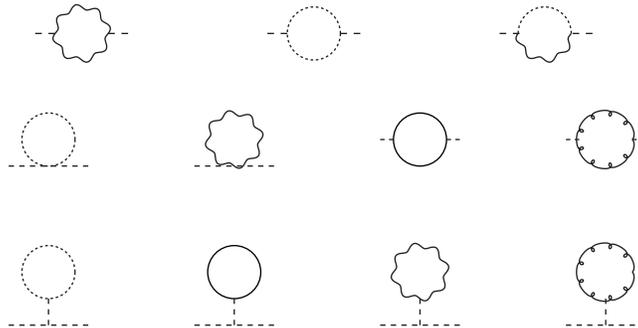
\\{\sl Fig. 3. Diagrams contributing to scalar self energy.}
\end{center}

Although calculations of these diagrams are given in the literature in order to make our point we will also give a detailed account here for our particular case of interest(massless Higgs and absence of trilinear coupling) and then compare with other results.
Let us give the various Higgs gauge fermion couplings:
\begin{eqnarray}
&&hW^{+}W^{-}\rightarrow\frac{2vm_W^2}{v^2}=\frac{vg^2}{2}
\nonumber\\
&&hZZ\rightarrow\frac{2vm_Z^2}{2v^2}=\frac{v(g^2+g'^2)}{2}
\nonumber\\
&&hhW^{+}W^{-}\rightarrow \frac{2m_W^2}{v^2}=\frac{g^2}{2}
\nonumber\\
&&hhZZ\rightarrow\frac{2m_Z^2}{2}=\frac{g^2+g'^2}{2}
\nonumber\\
&&ht{\bar t}\rightarrow \frac{m_t}{v}
\label{not67895}
\end{eqnarray}

Thus the first diagram and the fifth in Fig. 3 add up to:
\begin{eqnarray}
&&\Pi^{WW}=-i\frac{1}{2}g^2m_W^2\frac{1}{(2\pi)^4}\int dk^4\frac{1}{(k^2-m_W^2)^2}[g_{\mu\nu}-\frac{k_{\mu}k_{\nu}}{m_W^2}][g_{\mu\nu}-\frac{k_{\mu}k_{\nu}}{k^2}]+
\nonumber\\
&&+3i\frac{g^2}{4}\frac{1}{(2\pi)^4}\int d^4k \frac{1}{k^2-m_W^2}
\nonumber\\
&&=-i\frac{g^2}{2}m_W^2\frac{1}{(2\pi)^4}\int dk^4 \frac{3}{(k^2-m_W^2)^2}+3i\frac{g^2}{4}\frac{1}{(2\pi)^4}\int d^4k \frac{1}{k^2-m_W^2}
\label{one34}
\end{eqnarray}

This needs to be regularized but first let us extract the divergences from all the diagrams:
\begin{eqnarray}
&&\Pi^{ZZ}=-i\frac{g^2m_Z^2}{2\cos{\theta_W}}\frac{3}{(2\pi)^4}\int d^4k \frac{3}{(k^2-m_Z^2)^2}
\nonumber\\
&&+3i\frac{g^2}{4\cos^2{\theta_W}}\frac{1}{(2\pi)^4}\int d^4k \frac{1}{k^2-m_Z^2}
\label{one35}
\end{eqnarray}

The diagram with the fermion loop (we will consider here only the top quark as being the most significant) reads:
\begin{eqnarray}
&&\Pi^{t {\bar t}}=-i\frac{1}{(2\pi)^4}\frac{m_t^2}{v^2}\int d^4 k \frac{4k^2-m_t^2}{(k^2-m_t^2)^2}=
\nonumber\\
&&-\frac{m_t^2}{v^2} d^4 k [4\frac{k^2-m_t^2}{(k^2-m_t^2)^2}+3\frac{m_t^2}{(k^2-m_t)^2}]
\label{one5676}
\end{eqnarray}

Third diagram has the expression:
\begin{eqnarray}
\Pi^{hh}= i\frac{1}{(2\pi)^4}12\lambda\int d^4k \frac{1}{k^2-m_h^2}
\label{one768}
\end{eqnarray}

In general the tadpole diagrams contribute to the scalar self energy. However they do not contribute when there is no scalar cubic interaction. This is the case for the Coleman Weinberg model and this is the case here.
We use Pauli-Villars regularization procedure for the above diagrams. Thus we make:
\begin{eqnarray}
&&\int d^4 k\frac{1}{k^2-m_x^2}\rightarrow\int d^4 k\frac{1}{k^2-m_x^2}-\frac{1}{k^2-\Lambda^2}=
\frac{-(2\pi)^4i}{(4\pi)^2}[\Lambda^2-m_x^2\ln\frac{\Lambda^2}{m_x^2}]
\nonumber\\
&&\int d^4 k\frac{1}{(k^2-m_x^2)^2}\rightarrow\int d^4 k \frac{1}{k^2-m_x^2}-\frac{1}{(k^2-\Lambda^2)^2}=
\frac{(2\pi)^4i}{(4\pi)^2}\ln\frac{m_x^2}{\Lambda^2}
\label{reg45}
\end{eqnarray}

Now let us collect the whole self energy.
\begin{eqnarray}
&&\Pi=2\Pi^{WW}+\Pi^{ZZ}+\Pi^{t{\bar t}}+\Pi^{hh}=
\frac{1}{(4\pi)^2}g^2m_W^2\ln\frac{m_W^2}{\Lambda^2}+
\frac{1}{(4\pi)^2}\frac{g^2}{4}[\Lambda^2-m_W^2\ln{\frac{\Lambda^2}{m_W^2}}]
\nonumber\\
&&\frac{1}{(4\pi)^2}\frac{g^2 m_Z^2}{\cos^2{\theta_W}}\ln {\frac{m_Z^2}{\Lambda^2}}+
\frac{1}{(4\pi)^2}\frac{g^2}{4\cos^2{\theta_W}}[\Lambda^2-m_Z^2\ln{\frac{\Lambda^2}{m_Z^2}}]
\nonumber\\
&&\frac{-1}{(4\pi)^2}\frac{m_t^2}{v^2}[12\Lambda^2-3m_t^2\ln{\frac{\Lambda^2}{m_t^2}}]+
\nonumber\\
&& \frac{1}{(4\pi)^2}12\lambda[\Lambda^2-m_h^2\ln{\frac{\Lambda^2}{m_h^2}}]=
\nonumber\\
&&[\frac{3}{16\Pi^2}\Lambda^2[2m_W^2+m_Z^2+6\lambda-4m_t^2]+\frac{3}{16\pi^2}
[6\frac{g^4}{16}\frac{1}{v^2}\ln\frac{m_W^2}{\Lambda^2}+\frac{3(g^2+g'^2)^2}{16}\frac{1}{v^2}\ln\frac{m_Z^2}{\Lambda^2}-
\frac{g_Y^4}{4}\ln\frac{m_t^2}{\Lambda^2}]
\label{fin7676543}
\end{eqnarray}

First note that the result agrees with that in \cite{Ma}.
We give bellow the result for the second derivative of the potential at the minimum $\langle\Phi\rangle=f$.
\begin{eqnarray}
&&\frac{\partial^2V}{\partial \phi^2}=\frac{24}{32\pi^2}a_1\Lambda^2+\frac{3}{32\pi^2}[\frac{3}{4}g^4f^2\ln(\frac{g^2f^2}{4\Lambda^2})+
\frac{1}{2}g^4f^2+\Lambda^2g^2]
 \nonumber\\
&&\frac{3}{64\pi^2}[\frac{3}{4}(g^2+g'^2)^2f^2\ln(\frac{(g^2+g'^2)f^2}{4\Lambda^2})+
\frac{1}{2}(g^2+g'^2)^2f^2+\Lambda^2(g^2+g'^2)]
\nonumber\\
&&-\frac{3}{16\pi^2}[3g_Y^4f^2\ln(\frac{g_Y^2f^2}{2\Lambda^2})+2g_Y^4f^2+2\Lambda^2g_Y^2]
\label{secderi}
\end{eqnarray}
Then as we compare the above result with Eq (\ref{fin7676543}) we see that they actually disagree. In \cite{CW} Coleman Weinberg stated that the scalar self energy should be given by the second derivative of the effective potential at the true minimum. Later Weinberg showed in \cite{W} that this is true only if one works in the Landau gauge. He also proved this result for the Goldstone bosons. The reason why our two expressions disagree stems from the method of eliminating the divergences for the two cases. If one first derives with respect to the field in the integral for the effective potential and then integrates, one obtains the same result for the two expressions. This is pure technicality. Moreover if one considers the next term in the Coleman Weinberg expansion (see Chapter VII in \cite{W}) and one reinforces the vanishing of the mass at one loop, one obtains that the effective potential has a minimum at the tree level minimum value. Thus the effective minimum remains at the same point. This is perfectly true in our case.

As we do not wish to consider renormalization at this point we will consider the Eq (\ref{fin7676543}) as the reliable answer and work with it.

\section{Quadratic divergences}

In 1981 Veltman  \cite{Veltman}proposed  a solution to the
tuning problem of the Higgs boson by demanding that the
quadratic term in the effective potential be zero. This can cancel
the one loop contributions but in order to be consistent we would
have to put to zero the coefficients of $\Lambda^2$ in all order of
perturbation theory. While this is not possible in general there are
solutions to circumvent this problem \cite{Murayama}. The predicted
Higgs mass for this case is 317$\pm$11 GeV.

Again we are starting with the usual Higgs potential:
\begin{equation}
U(\phi)=\frac{1}{2}\mu^2(\Phi^{\dagger}\Phi)+\frac{1}{4}\lambda
(\Phi^{\dagger}\Phi)^2\label{thepotential}
\end{equation}

 The correction to the mass reads from
 Eq (\ref{onelooppotential1})-(\ref{results3})
and we get \cite {Murayama} in terms of the cut-off:

\begin{equation}
\mu_R^2=\mu^2+\frac{3}{32\pi^2
v^2}\Lambda^2[2m_W^2+m_Z^2+m_h^2-4m_t^2] \label{masstotal}
\end{equation}

Here all the masses are calculated at the minimum of the potential
$\phi=v$. Thus for
an arbitrary vev of the field $\langle\Phi\rangle$ the mass of the
Higgs is $m_h^2=\mu^2+3\lambda \langle\Phi\rangle^2$ while the mass
of the Goldstone bosons is $m_G^2\mu^2+\lambda\langle\Phi\rangle^2$.
At the minimum $\langle\Phi\rangle=v$ the masses become
$
m_h^2=2\lambda v^2=-2\mu^2$ and $m_G^2=0$.

The question is: how would the Veltman condition read for a general form of the Higgs potential?
If we look at Eq (\ref{result1})-(\ref{results3})we can see immediately that all it is to do is the derive those expressions twice with respect to the Higgs. Then
\begin{equation}
\mu_R^2=\mu^2+\frac{3}{32\pi^2
v^2}\Lambda^2[2m_W^2+m_Z^2+\frac{U''''v^2}{3}-4m_t^2]
\label{masstotal2}
\end{equation}

 It should be evident by now that by changing the form of the potential we can improve drastically the Veltman condition. In particular by introducing a dilaton with an associate scale f, v will get replaced with f in the above condition thus opening the possibility of manipulating it.

\section{Keeping it simple}

First we will require by hand the cancelation of the quadratic divergences at the minimum of the potential (the Veltman condition). Note that this leads to a relation between the dilaton self coupling and the scale f but does not imply a nonzero mass:

\begin{equation}
\frac{3}{32\pi^2}[8a_1+ g^2+\frac{1}{2}(g^2+g'^2)-4g_Y^2]=0
\label{cond99}
\end{equation}

With the scaled coupling constants given in (\ref{scaling}) and the known expression for the masses this amounts to:
\begin{equation}
16 a_1 f^2+2m_W^2+m_Z^2-4m_t^2=0
\label{cond100}
\end{equation}
We will not give a second thought to Eq (\ref{cond100}) as it can be easily realized and $a_1$ can be made as small as desired as long as the scale f is high enough.
Note that as we cancel the quadratic divergences the terms that are left in Eq(\ref{secderi}) are all suppressed by $\frac{v^2}{f^2}$.
If we require the cancelation of both terms and solve for the cut-off we see that it is actually at the electroweak scale. Thus the model survives very well and it is perfectly consistent with the standard model couplings, the drawback being the limits imposed by LEP. As these are very stringent we are left with the approximate cancelation of the logarithmic term due to the suppression factor.
 Note that this was first proposed by Ma in \cite{Ma}where he requires the simultaneous cancelation of the quadratic divergences and of the logarithmic ones thus rendering the one loop correction to the Higgs mass finite. Unfortunately this leads to the wrong top quark mass making this attempt unsuccessful.
 However there is nothing to prevent us instead of canceling separately the quadratic divergences and the logarithmic ones to require that the whole correction is zero and solve for $a_1$. In this way we obtain results compatible with very high f, at least of order $10^7$ GeV.

Assume this limiting case:
we break the symmetry and calculate the one loop potential for f identified with the  cut-off scale. Then we make $f\rightarrow\infty$. We still can obtain the right masses for the fermions and for the gauge bosons but the Higgs boson decouples completely. Thus particle have masses but do not couple with the Higgs which remains massless also at one loop level.

\section{Phenomenology}

 First let us briefly mention that a zero mass Higgs is not affected by the theoretical constraints coming form triviality,unitarity and vacuum stability since these put only upper bounds on the mass.

 LEP experiments were looking for  a light scalar mainly through the decay (Bjorken process)
 \begin{equation}
 e^+e^-\rightarrow ZH
 \label{decay1}
 \end{equation}

 insensitive to the decay mode of the scalar. Thus they extract upper bounds to the following cross section or k:
 \begin{equation}
 \sigma_{S^0Z^0}=k\sigma_{H_{SM}Z^0}
 \label{crossk}
 \end{equation}

 Of course the cross section becomes larger as $\lambda^{\frac{1}{2}}[\lambda+12sm_Z^2]$ where $\lambda =(s-m_h^2-m_Z^2)^2
 -4m_h^2m_Z^2$ increases so it attain its peak for $m_S=0$. But in our model the Weinberg angle is that of the SM  but the coupling constants scale as $\frac{g}{f}$. Thus any increase in the cross section can be highly suppressed by the scale factor.

 For the electroweak precision data we will mention only the contribution to the Z decay width through the branching ratio:
 \begin{eqnarray}
 \frac{BR(Z\rightarrow H f f)}{BR(Z\rightarrow f f)}(m_H=0)&=&\frac{g^2}{192\pi^2cos^2\theta_W}
 [(6-\frac{\Gamma_Z^2}{2m_Z^2})ln(\frac{\Gamma_Z^2+m_Z^2}{\Gamma_Z^2})
 \nonumber\\
 &&+\frac{12\Gamma_Z}{m_Z}tan^{-1}(\frac{m_Z}{\Gamma_Z}-\frac{23}{2}).
 \label{branch}
 \end{eqnarray}

 Note that the highest contribution is for $m_H=0$ and even for this case of the SM this is pretty low $\approx 10^{-2}$. For our case it is again suppressed by the scale factor so it gets even smaller.

Among the usual Higgs decays only the one loop ones to two photons or two gluons are kinematically allowed so there is  avery high probability that the Higgs passes through detector undetected. Let us analyze the main source of Higgs production at hadron colliders, the gluon fusion:
\begin{equation}
\frac{d\sigma_h}{dy}=\frac{\alpha_s^2g^2}{1024\pi m_W^2}\frac{2}{3}g_A(x_a,m_h^2)g_B(x_b,m_h^2)
\label{gluonfusion}
\end{equation}

where y is the rapidity and for the case of a massless Higgs $x_a=0$ and $x_b=0$. Then using the resources given in
\cite{Kane2}we can estimate the actual value which is of order $0.3$
pb. In terms of the standard model cross section of this type this would correspond to a SM Higgs in the range of $\approx 100-150$ GeV.

Other important limits on Higgs boson mass come from nuclear experiments. While the vast majority of them have a range of masses that are prohibited thus allowing a massless Higgs boson there are two that need to be considered. One of them is the lower limit on the calculated branching ratio for the decay $K^{\pm}\rightarrow\pi^{\pm}\phi^0$ which associated with a massless Higgs and the higher experimental limits gives a contradiction. The escape is easy for our case since by increasing the scale f we can make the lower limit as low as we want and thus at least lower than the experimental one. The second one is coming from the
the limits on $a_N$ in the neutron charge form factor $G_e^N(t)=a_Nt$. This assumes \cite{Kane2}that the electron-neutron Coulomb interaction can be written as:
\begin{equation}
V_{eN}\rightarrow(\frac{e^2G^N_e(t)}{t}-\frac{g_{hee}g_{hNN}}{t-m_h^2})
\label{elnint}
\end{equation}

 and eliminates categorically the possibility of  completely massless Higgs, leaving open the possibility of Higgs of a mass however small.  A possibility of avoiding this constraint is to couple the quarks and the leptons with different fields thus avoiding the possibility of modifying the electromagnetic interaction.
 There are two possibilities: one for example is to couple the quarks with $\chi$ while the leptons couple with $\sigma=f\ln{\frac{\chi}{f}}$. In this way however we need a different mechanism to give masses to the charged leptons. Another possibility is to couple the quarks with the full Higgs matrix and the charged leptons only with the Goldstone bosons.
As an aside let us give suggestion of how the Higgs matrix might couple with the fermion doublets (see for example\cite{hamed}).
\begin{eqnarray}
\left[
\begin{array}{cc}
{\bar \nu}_L & {\bar e}_L
\end{array}
\right]
f\exp [I \frac{1}{f}\left[
\begin{array}{ccc}
\pi_0-I\sigma&\pi_1-I\pi_2\\
\pi_1+I\pi_2&-\pi_0-I\sigma
\end{array}
\right]
\left[
\begin{array}{cc}
0\\
1
\end{array}
\right]
e_R
\label{couplfer4563}
\end{eqnarray}

It can be checked that with the double invariance of the Higgs matrix that the above expressions is still invariant under an $SU(2)$ with the correct quantum numbers.

\section{Conclusions}
    In this paper we assume, contrary to the common knowledge(wisdom) that light scalars do not exist in nature, that the Higgs boson is actually completely massless.
     We propose one simple scenario where this is realized at tree level and can be extended at one loop level despite the fact that the gauge symmetry is broken and the particles get masses.
     Of course it is well known that gauge invariance, Lorenz invariance and renormalizability coexist only with the usual form of the Higgs potential. Thus we have to give up one of the above criteria, so we discard renormalizability which is natural for the type of dilaton potential we propose.

    We consider that in the context of a scale invariant theory with the couplings sensibly diminished with respect to those of the standard model it is possible for massless particle to have escaped detection at the various high energy colliders. Then we show how various  experimental constraints, mostly of nuclear nature can be avoided.
\section*{Acknowledgments} \vskip -.5cm
We are happy to thank A. D. Polosa, J. Schechter for very useful comments on the manuscript and A. Fariborz for fruitful discussions.


\begin{thebibliography}{15}


\bibitem{Kane1} Gordon L Kane,"Perspectives on Higgs Physics",
 World Scientific Publishing Co. Pte. Ltd. (1993).
 \bibitem{Kane2} J. F. Gunion, H. H. Haber, G. Kane and S. Dawson,
"The Higgs Hunter's guide", Perseus Publishing (1990).
\bibitem{Skiba1} W. D. Goldberger, B. Grinstein and W. Skiba, arXiv:0708.1463(hep-ph).
\bibitem{Skiba2} J. Fan, W. D. Goldberger, A. Ross and W. Skiba, arXiv:0803.2040(hep-ph).
\bibitem{JS} R. Jora, S. Moussa, S. Nasri, J Schechter and M. N. Shahid, Int. J. Mod. Phys.423:5159-5172,2008,
 arXiv:0805.0293.
\bibitem{Sher1} M. Sher, Phys. Rep. {\bf 179},Nos. 5 and 6, 273-418
(1989).
\bibitem{MV} M. E. Machacek and M. T. Vaughn, Nucl Phys.{\bf B 222},
83 (1983), Nucl. Phys. {\bf B 236} 221 (1984), Nucl. Phys. {\bf B
249} 70 (1985).
\bibitem{CW} S. Coleman and E. Weinberg, Phys. Rev D {\bf 7} 1888
(1973).
\bibitem{W} S. Weinberg, Phys. Rev. D {\bf 7} 2887 (1973).
\bibitem{Veltman} M. Veltman, Acta Phys. Polon. {\bf B 12},
437 (1981).
\bibitem{Murayama} C. Kolda and H. Murayama, arXiv:hep-ph/000317.
\bibitem{TA} T. Appelquist, Phys. Rev D {\bf 22} 22 (1980).
\bibitem{Ma} E. Ma, arXiv;hep-ph/920922.
\bibitem{hamed} M Schmaltz and D. Tucker-Smith, arXiv:hep-ph/0502182(and also the references there).
\end{thebibliography}
\end{document}